\documentstyle[preprint,revtex]{aps}
\begin{document}
\draft
\begin{title}
  ATOMIC PARITY NONCONSERVATION:\\
 ELECTROWEAK PARAMETERS AND NUCLEAR STRUCTURE
\end{title}
\author{S.J. Pollock}
\begin{instit}
Institute for Nuclear Theory, University of Washington,
Seattle, WA 98195
\end{instit}
\author{E. N. Fortson and L. Wilets}
\begin{instit}
 Department of Physics, University of Washington,
Seattle, WA 98195
\end{instit}
\receipt{May 12 1992}
\begin{abstract}
There have been suggestions to measure atomic parity nonconservation
(PNC) along an isotopic chain, by taking ratios of observables in order
to cancel complicated atomic structure effects.  Precise atomic PNC
measurements could make a significant contribution to tests of the
Standard Model at the level of one loop radiative corrections. However,
the results also depend upon certain features of nuclear structure,
such as the spatial distribution of neutrons in the nucleus.  To
examine the sensitivity to nuclear structure, we consider the case of
Pb isotopes using various recent relativistic and non-relativistic
nuclear model calculations.  Contributions from nucleon internal weak
structure are included, but found to be fairly negligible.  The spread
among present models in predicted sizes of nuclear structure effects
may preclude using Pb isotope ratios to test the Standard Model at
better than a one percent level, unless there are adequate independent
tests of the nuclear models by various alternative strong and
electroweak nuclear probes.  On the other hand, sufficiently accurate
atomic PNC experiments would provide a unique method to measure neutron
distributions in heavy nuclei.
\end{abstract}
\narrowtext
\section{INTRODUCTION}
Parity nonconservation (PNC) in atoms arises from the
electroweak interaction between the electrons and nucleons, primarily
due to exchange of the neutral gauge boson, $Z_0$.  The dominant
contribution in heavy atoms comes from the coupling of the axial
electronic current to the vector nucleon current.  Because the vector
currents are conserved, atomic PNC essentially measures the electroweak
coupling to the elementary quarks, bypassing many of the difficulties
of hadronic physics. Thus in principle atomic experiments can measure
certain key electroweak parameters quite accurately, and also help
probe for new physics beyond the presently successful Standard Model of
the electroweak interactions.

In fact, there remains much to be learned about the Standard
Model, including the masses of the top quark and the predicted
Higgs boson(s), and whether there are additional generations of quarks
and leptons.  In addition the Standard Model faces the well-known gauge
hierarchy problem, and it is certainly possible that electroweak
measurements may reveal something totally new, such as technicolor or
supersymmetric particles.  Accurate measurements of PNC in atomic
cesium already play an important role in addressing such questions.

Two major issues affect the interpretation of atomic experiments and
will become more crucial as experimental accuracy improves, namely the
small but not negligible effects of nuclear size and
structure,\cite{fpw} and the reliability of the atomic theory of heavy
atoms\cite{bjs,dfs}. Atomic theory, the source of the largest
uncertainty, has received a great deal of attention leading to
increasingly precise calculations of PNC for a number of elements.
Cesium, in particular, is now believed to be understood at the 1\%
level.

To advance further may require canceling out all uncertainties of the
atomic theory by comparing PNC measurements on different isotopes of
the same element.  Such experiments in fact have been
proposed\cite{dfk} using strings of isotopes of such elements as
Cs, Dy, and Pb.

As we discussed in a previous note,\cite{fpw} hereafter
referred to as I, it then becomes important to find the level at which
nuclear structure interferes with interpreting atomic PNC purely in
terms of particle theory.  The wave function of the atomic electrons
varies over the dimensions of the nucleus, causing the net electroweak
interaction with the nucleons to depend on the spatial distribution of
both the protons and neutrons.  As demonstrated in I, the PNC
observable is (for $\sin^2\theta_W\approx 1/4$ and $R_n\approx R_p$)
very roughly proportional to
\begin{equation}
1-{\textstyle {3\over 70}}
   (Z\alpha)^2\big[1+5\,R_n^2/R_p^2 \big] + \cdots , \eqnum{1.1}
\end{equation}
where $R_n$ and $R_p$ are the equivalent rms radii for the nuclear
distribution of  neutrons and protons. The proton, or rather the
nuclear {\it charge}, distribution is well known from electric probes:
electron and muon scattering, optical isotope shifts, muonic atoms,
etc.  The extraction of the neutron distribution, however, is quite
model-dependent and difficult to determine to the same high accuracy.

On the one hand, the neutron distribution is needed in order to extract
the weak parameters in heavy atom experiments.  On the other hand, to
the extent the weak parameters are known, the experiments provide a
method of {\it measuring} the changes in the neutron distribution,
primarily the rms radius.  Thus atomic experiments on isotopes of heavy
atoms may provide a unique opportunity to test nuclear model
calculations.

In this paper we explore the nuclear structure issues
extensively. We have utilized several recent detailed nuclear structure
calculations from various authors, in order to quantitatively estimate
the nuclear model-dependent corrections to atomic PNC in the isotopes
of $_{82}$Pb, an element of interest experimentally.  Our major
conclusions are:

(1) For single isotope measurements on $^{208}$Pb, $(Z\alpha)$ is
sufficiently large that the effects of nuclear structure on atomic PNC
cannot be neglected. The {\it uncertainties} due to neutron
distributions  appear to be less significant for the extraction of
electroweak parameters than the current uncertainties  due to atomic
structure. However, unless  the {\it change} in neutron distributions
along an isotopic chain can be better predicted (or independently
measured, e.g. via parity violating e$^-$ scattering), PNC ratios of
$_{82}$Pb isotopes will not be able to provide an extraction of the
weak mixing angle to better than a one percent level.

(2)  In lighter nuclei (including the important case of $_{55}$Cs),
$Z\alpha$ is sufficiently small that uncertainties in nuclear structure
can probably be safely ignored compared to current uncertainties in
atomic theory, when using a single isotope. Calculations for the
non-magic, odd-$Z$, Cs nuclei pose additional difficulties, and further
investigation is still required to determine how accurately the
Standard Model could be tested when using ratios of isotopes.

(3)  Atomic PNC experiments provide perhaps one of the cleanest
opportunities yet available to study the nuclear neutron skin.  The
situation is similar to atomic isotope shifts which have provided
precise measurements of the ratios of changes in the nuclear rms charge
radius in strings of isotopes.  Here we have a weak probe of the
neutron distribution which is free of the gross uncertainties
associated with strongly interacting probes.  At the level of precision
that the Standard Model is known, this yields another testing ground
for nuclear models.

We note that alternative weak probes, such as parity-violating
intermediate-energy $e^-$-nucleus cross section asymmetries, are also
sensitive to the neutron distributions. In combination with atomic PNC,
these may help simplify the separation of nuclear structure effects
from electroweak radiative corrections.

The paper is organized as follows:  In section II we sketch briefly the
relevant parts of electroweak theory.  In section III we review the
simple analytical model we presented in reference (1); this provides a
convenient framework for discussing the effects of nuclear structure on
atomic PNC in terms of the rms radii of the proton and neutron
distributions in the nucleus.  In section IV we discuss the electroweak
nucleon form factors.  The intrinsic electroweak structure does begin
to contribute at the level we are interested in, but the {\it
uncertainties} in this structure (due to strangeness admixtures, etc.)
should have a negligible effect on the total PNC amplitude of the
nucleus.  In section V, we examine the key ingredients of currently
available theoretical models for heavy nuclei, including both
non-relativistic and relativistic formalisms.  We consider the
reliability of these models, and discuss the need for calculations
which include correlations beyond the Hartree-Fock level. We also
discuss alternative experimental means to measure the desired neutron
distributions.  In section VI we discuss the relevant Standard Model
parameters, and the accuracy desired in their extraction from atomic
parity violation.  We derive the propagation of error from nuclear
model uncertainties to electroweak parameters, focusing on isotopes of
lead and (to a lesser extent) cesium.  These two elements are of
current experimental interest, and are representative of the very heavy
and moderately heavy regions of the periodic table.  In section VII we
discuss our numerical results, using various existing Hartree-Fock
nuclear calculations, summarize, and discuss the conclusions from the
previous sections.

\section{  PNC IN THE STANDARD MODEL}
\subsection{ Theoretical Considerations}

Because the $Z$-boson is massive (91.16$\pm$.03 GeV), the
quark-electron interaction due to $Z$-exchange may be taken to be of
zero range compared to atomic or nuclear dimensions.  What we observe
in atoms is the electron interaction with {\it nucleons}, not
individual quarks.  Nucleons are, of course, composite structures
consisting each of three quarks net, but also $\bar q q$ pairs as well
as gluons.  We make the assumption here, which we justify in Sec. IV,
that we can neglect the internal nucleon structure and simply add the
point coupling of the 3 quarks to obtain the net nucleon weak
coupling.  The PNC part of the nucleon-electron interaction can be
written in terms of axial and vector currents
\begin{equation}
H_{PNC}=V_N\times A_e+A_N\times V_e\,.            \eqnum{2.1}
\end{equation}
If, in addition to neglecting internal nucleon structure, we treat the
nucleons nonrelativistically (a very good approximation), we have
\FL
\begin{eqnarray}
H_{PNC}={G_F\over
\sqrt{2}}\ \sum_{eB}\ \biggl[&& C_{1B}\int\psi^\dagger_B\psi_B
    \psi^\dagger_e   \gamma^5\psi_e  d^3\!r \nonumber\\
    &&\ +\ C_{2B}\int\psi^\dagger_B\vec \sigma_B\psi_B\cdot\psi^\dagger_e
    \vec \alpha\psi_e d^3\!r\biggr]\,,\nonumber\\
&&      \eqnum{2.2a}
\end{eqnarray}
where $B$ stands for $n$ (neutron) or $p$ (proton) and
\begin{eqnarray}
C_{1p}\ =&&\ {\textstyle {1\over 2}}(1-4 \sin^2\theta_W),\nonumber\\
C_{2p}\ =&&\ {\textstyle {1\over 2}}
(1-4 \sin^2\theta_W)g_{\scriptstyle A},
				\nonumber\\
C_{1n}\ =&&\ -{\textstyle {1\over 2}},\eqnum{2.2b}\\
C_{2n}\ =&&\ -{\textstyle {1\over 2}}
		(1-4 \sin^2\theta_W)g_{\scriptstyle A}.
				\nonumber
\end{eqnarray}
These expressions assume tree-level Standard Model couplings. In section VI,
we discuss the important effects of radiative loop corrections.

The first term in eqn. (2.2a) grows coherently with nucleon numbers $N$
and $Z$.  The second term, together with the anapole
moment\cite{nmw} term (which also depends upon $\vec
\sigma_B\cdot\vec \alpha$), amounts to at most a few percent of the
first term in heavy atoms, and furthermore sums to zero when all hfs
sublevels are combined, since all directions of $\vec \sigma_B$ are
then weighted equally.  Thus in this paper we will consider the first
term only.  The effective interaction is
\FL
\begin{eqnarray}
H_{PNC,1}={G_F\over 2\sqrt{2}}\int&& \left[-N\rho_n(\vec r\,)+
      	Z(1-4\sin^2\theta_W)\rho_p(\vec r\,)\right]\nonumber\\
&&\quad \times\   \psi^\dagger_e \gamma^5\psi_e\, d^3\!r\,,\eqnum{2.3}
\end{eqnarray}
where here the $\rho_n$ and $\rho_p$ are normalized to unity.  The
neutron and proton densities include a folding with the weak form
factors (see Sec. IV).

We need the spatial variation of $\psi^\dagger_e\gamma^5\psi_e$ over
the nucleus, its normalization, and its dependence on nuclear
structure. PNC effects are dominated by $s_{1/2}$-electrons
($\kappa=-1$) coupled to $p_{1/2}$-electrons ($\kappa=+1$). We define
\begin{equation}
\rho_5(r)\equiv \psi^\dagger_p(\vec r\,)
\gamma^5\psi_s(\vec r\,)\,,	\eqnum{2.4}
\end{equation}
which turns out to depend only on the magnitude of $\vec r$.
$\rho_5(r)$ can be factored conveniently as follows:
\begin{equation}
\rho_5(r) =C(Z)\,{\cal N}(Z,R)\,f(r)\,,	\eqnum{2.5}
\end{equation}
where $C(Z)$ contains all atomic structure effects for a point
nucleus including many-body correlations; ${\cal N} \equiv
\psi^\dagger_p(0) \gamma^5\psi_s(0)$ is the normalization factor for a
single electron;  $f(r)$ contains the spatial variation and is
normalized to $f(0)=1.$

Because the electric potential is very strong
near the nucleus, we can safely neglect atomic binding energies in
$f(r)$.  In Pb, for example,  the potential at the nuclear surface is
about 15 MeV compared with valence electron binding energies of a few
eV. In addition, to a very good approximation,\cite{fpw}
\begin{equation}
{\cal N}= R^{-\gamma}\,, \eqnum{2.6}
\end{equation}
where $\gamma=2\Big[1-\sqrt{1-(Z\alpha)^2}\Big]$ and
$R$, often called the {\it equivalent charge radius,} is given by
\begin{equation}
 R=\left[{\textstyle {5\over 3}}
<\! r^2\!>_{charge}\right]^{1/2}. \eqnum{2.7}
\end{equation}
We are not interested in the absolute value of ${\cal N}$, but only its
variation with nuclear structure.  Observable PNC effects are
proportional to the matrix element between two atom\-ic states $i$ and
$j$
\FL
\begin{eqnarray}
<\! i|H_{PNC,1}|j\!>=&&
{G_F\over 2\sqrt{2}}\,C_{ij}(Z)\,{\cal N}\nonumber\\
&&\ \times\  \big[-N\,q_n+Z(1-4\,\sin^2\theta_W)q_p\big]\,.\nonumber\\
&&\eqnum{2.8}
\end{eqnarray}
As mentioned above, this is modified by radiative
corrections which we discuss in some detail in section VI.
Effects of {\it nuclear} structure on PNC  are contained in ${\cal N}$
and the two quantities
\begin{eqnarray}
q_n\ &=&\ \int \rho_n(r)\,f(r)\,d^3\!r\,,	\eqnum{2.9\,a}\\
q_p\ &=&\ \int \rho_p(r)\,f(r)\,d^3\!r\,.	\eqnum{2.9\,b}
\end{eqnarray}
We note that $1-4\,\sin^2\theta_W$ is a small number; from high energy
experiments, $\sin^2\theta_W=0.230\pm .004.$  The value of
$\sin^2\theta_W$ can also be deduced from atomic experiments with an
accuracy that will be limited in part by nuclear structure effects, as
we discuss in sections VI and VII.

The proton (charge) nuclear form factors needed for $q_p$ and ${\cal
N}$ are generally well known from measurements of the charge
distribution of nuclei close to the stable valley and many unstable
nuclides as well.  Neutron nuclear form factors are needed for $q_n$,
and are not well-determined experimentally, and statements about them
are quite model-dependent.  Neutron and proton distributions are often
taken to be proportional to each other, scaled by $N$ and $Z.$ However,
neutron-rich nuclei have larger neutron distributions than the protons
and the reverse is true for proton rich nuclei.  In an isotopic
sequence, the $A^{1/3}$ law is not followed for either the charge or
the neutron distributions separately.

\subsection{Experimental Considerations}

As first pointed out by Bouchiat and Bouchiat,\cite{bou} the effect of
$H_{PNC}$ in neutral atoms grows rapidly with atomic number $Z$,
approximately as $Z^3$.  Thus experimental interest has concentrated on
heavy atoms, namely $_{55}$Cs, $_{81}$Tl, $_{82}$Pb, and $_{83}$Bi.
(For some reviews, see reference \cite{rev}.) The measured quantity in
all experiments is the electric dipole amplitude ${\cal E}_{PNC}$
between two electronic states which, in the absence of $H_{PNC}$, would
have the same parity and hence would have no electric dipole amplitude
connecting them.  Denoting the initial and final states by $i$ and $f$,
we can write:
\FL
\begin{eqnarray}
{\cal E}_{PNC}=\sum_n&&\biggl[
	 {<f\vert \hat {\bf E}_1\vert n> <n\vert H_{PNC}\vert i>\over
	 		W_i-W_n}\   \nonumber\\
	&&\quad\quad +\  {<f\vert H_{PNC}\vert n>
			<n\vert\hat {\bf E}_1\vert i>\over
			 W_f-W_n}\biggr], \eqnum{2.10}
\end{eqnarray}
where the first and second terms give the mixing due to $H_{PNC}$ of
opposite parity states into the initial and final states respectively.
$W$ is the energy of the atomic states, and $\hat {\bf E}_1 \equiv
-\sum_j e{\bf r}_j$ is the electric dipole charge operator.  The
magnitude of ${\cal E}_{PNC}$ is of order $10^{-9} ea_0$ for the
heaviest atoms of interest.

Two experimental techniques have evolved for measuring ${\cal
E}_{PNC}$.  One involves applying an external static electric field
which, like $H_{PNC}$, mixes in opposite parity states and creates an
electric dipole amplitude between the states $i$ and $f$.  The
interference between this Stark amplitude and ${\cal E}_{PNC}$ leads to
a parity-violating signature in the optical transition from $i$ to $f$
in which the sign of the interference term reverses with the sense of
circular polarization of the incident light, and with other vectors
specifying the handedness of the experimental arrangement.  The other
technique uses no external fields, but instead exploits the
interference between ${\cal E}_{PNC}$ and the magnetic dipole ($M1$)
amplitude between the same two states.  This interference causes
parity-violating optical rotation, i.e. a rotation of the plane of
polarized light passing through the atomic vapor at wavelengths near
the magnetic dipole absorption line.

The Stark interference technique has been used in the all-important
measurement of PNC in Cs on the highly forbidden $6S_{1\over 2}$ -
$7S_{1\over 2}$ $M1$ absorption line at 532 nm, and in the measurement
on the $6P_{1\over 2}$ - $7P_{1\over 2}$ $M1$ transition in Tl.  The
optical rotation technique has been applied to the allowed $M1$
absorption lines at 876 nm and 648 nm in Bi, and to the similar 1278 nm
and 1283 nm lines in Tl and Pb respectively; all of which involve
transitions among low-lying configurations of $6p$ electrons, for
example $6P_{1\over 2}$ - $6P_{3\over 2}$ in Tl.  Both techniques have
reached the 1 percent level of accuracy.

Among the elements studied thus far, Cs and Pb are the most likely
candidates for comparing different isotopes.  It may be possible in the
case of Cs to use optical atom traps to carry out measurements on a
long string of radioactive isotopes.  Measurements on Pb will probably
be restricted to stable or long-lived isotopes.  In either case,
achieving the level of accuracy discussed in this paper (a few percent
down to 1 percent in the isotopic {\it difference}), although possible
in principle, will be a challenging task in the next generation of
atomic PNC experiments.

\section{ A SIMPLE MODEL FOR THE NUCLEAR FORM FACTORS}

Given proton and neutron distribution functions, there is no difficulty
in calculating $q_p,\;q_n$ and the variation in ${\cal N}$.  In I, we
used a simple model to estimate the importance of nuclear structure on
PNC observables.  We review those results here.

      Consider a uniform nuclear charge distribution of
radius $R$. This charge produces an electric potential
\begin{equation}
V_c(r)=Ze^2\Bigg\{
{(-3+r^2/R^2)/2R\,,\qquad r<R\,,
\atop -1/r\,,\;\;
\quad\qquad\qquad\qquad       r>R\,.}	\eqnum{3.1}
\end{equation}
A power series for the Dirac wave function {\it inside} the nucleus
yields
\begin{eqnarray}
f(r)=1-{\textstyle {1\over 2}} (Z\alpha)^2&&\Bigl[
(r/R)^2 -{\textstyle {1\over 5}} (r/R)^4  +
	 {\textstyle {1\over 75}} (r/R)^6\Bigr]\nonumber\\
&& + \ {\cal O} (Z\alpha)^4 \,. 	\eqnum{3.2}
\end{eqnarray}
Again for the sake of simplicity here, we assume that, as for a uniform
distribution (for either $n$ or $p$), $<r^4>={\textstyle {3\over 7}} R^4$ and
$<r^6>={\textstyle {3\over 9}} R^6$
where $R^2\equiv {\textstyle {5\over3}} <r^2>$.
{}From (2.9a) we find (neglecting here any differences
between charge and proton radii)
\begin{equation}
q_p=1-0.260 (Z\alpha)^2 +
{\cal O} (Z\alpha)^4\,,   \eqnum{3.3\,a}
\end{equation}
which is insensitive to nuclear structure {\it to this order.}
{}From (2.9b), we find
\FL
\begin{eqnarray}
q_n&=&1-(Z\alpha)^2\left({3\over 10}
{R_n^2\over R_p^2}-{3\over 70}
{R_n^4\over R_p^4}+{1\over 450}
{R_n^6\over R_p^6}\right)+
{\cal O} (Z\alpha)^4\nonumber\\
&\approx& 1- (Z\alpha)^2
\left(0.038+0.221{R^2_n\over R^2_p}\right)
+{\cal O} (Z\alpha)^4    \,,     \eqnum{3.3\,b}
\end{eqnarray}
which does depend on the neutron form factor. Here we have introduced
equivalent neutron and proton radii of the form (2.7); the second form in
(3.3b) assumes that $(R_n/R_p)^2-1$ is small.

      In this section, we have made rough approximations in order to
illustrate the sensitivity of the results to moments of the neutron and
proton distributions.  For comparison with experiment, a more detailed
analysis is necessary, using actual solutions of the Dirac equation for
realistic charge distributions and the best available theoretical
neutron distributions.  This is done in Secs. VI and VII.

\section{INTRINSIC NUCLEON STRUCTURE EFFECTS}

The usual treatment of atomic PNC begins with an effective Hamiltonian
for the parity violating electron-nucleus interaction, as in equation
(2.2a), which involves normalized proton and neutron distributions

\begin{eqnarray}
Z \rho_p(\vec r\,)\ =& &\ \sum_p \langle \psi_p^\dagger (\vec r\,)
\psi_p(\vec r\,)\rangle,
      \eqnum{4.1a}\\
N \rho_n(\vec r\,)\ =& &\  \sum_n \langle \psi_n^\dagger (\vec r\,)
\psi_n(\vec r\,)\rangle,\eqnum{4.1b}
\end{eqnarray}
where $\psi_N^{(\dagger)}$ is a destruction (creation) operator for
nucleons, and the matrix elements are between nuclear ground states.
However, these formulae implicitly assume point-like nature for
nucleons, and thus the usual analysis makes no distinction between
weak, electromagnetic, or point nucleon distributions, aside from
overall charges.

Of course, nucleons do have an internal structure,  and this must be
properly folded into the above distributions.  The internal weak
structure is related to, but different from, the electromagnetic
structure, and can be calculated in the context of the Standard Model.
We demonstrate in this section that the known, electromagnetic
structure of nucleons yields a rather small overall effect on atomic
PNC calculations, but must be included when extremely high precision
results are required.

There has been considerable discussion in recent literature
\cite{emc,don,kap} concerning the possibility of nontrivial
strange quark matrix elements in the nucleon.  This could lead to a
sizable ``strangeness radius'' of the nucleon, which in turn would
modify the weak radius in a well defined way. We allow for this
possibility in our analysis, although such a strangeness contribution
to atomic PNC is likely to be quite negligible.

In the Standard Model, assuming in addition that strong SU(2) isospin
is a good symmetry for the nucleons, one can extract relations between
weak and electromagnetic form factors \cite{wei} which then
describe the internal nucleon structure:
\begin{eqnarray}
G_E^{weak,p}(q^2)\ &=&\ {\textstyle{1\over2}}(1-4\sin^2\theta_W)
G_E^{\gamma,p}(q^2)\nonumber\\
  &&\ \ -{\textstyle{1\over2}}\left(G_E^{\gamma,n}(q^2)+G_E^{s}(q^2)
      			\right),\eqnum{4.2a}\\
G_E^{weak,n}(q^2)\ &=&\ {\textstyle{1\over2}}(1-4\sin^2\theta_W)
G_E^{\gamma,n}(q^2)\nonumber\\
 &&\ \  -{\textstyle{1\over2}}\left(G_E^{\gamma,p}(q^2) +G_E^{s}(q^2)
			\right).\eqnum{4.2b}
\end{eqnarray}
Here, $G_E^{X,N}$ is the usual Sachs electric form factor for
a current operator $J_\mu^X$, where $X$ can represent weak, electromagnetic,
 or
specific quark flavor currents:
\FL
\begin{eqnarray}
<\!{p',N}|J_\mu^X|&&{p,N}\!>\  \nonumber\\
     &&\equiv \bar u(p')\biggl(F_1^{X,N}(q^2)\gamma_\mu\nonumber\\
      &&\qquad\qquad \ \ + i F_2^{X,N}\sigma_{\mu\nu}q^\nu /(2M) \biggr)u(p),
      				\eqnum{4.3a}\\
G_E^X(q^2)\ &&\equiv \
F_1^X(q^2)+(q^2/4M^2)F_2^X(q^2),\eqnum{4.3b}
\end{eqnarray}
and
\begin{eqnarray}
\ J_\mu^\gamma\ =& &\
 {\textstyle {2\over3}}(\bar u \gamma_\mu u)
 -{\textstyle {1\over3}}(\bar d\gamma_\mu d +
\bar s\gamma_\mu s)
\eqnum{4.4a}\\
      \equiv& &\ {\textstyle {2\over3}}(J_\mu^u)
 	-{\textstyle {1\over3}}(J_\mu^d + J_\mu^s),
\eqnum{4.4b}\\
\ J_\mu^{weak}\ =& &\
         ({\textstyle {1\over2}}-{\textstyle
{4\over 3}}\sin^2\theta_W)(J_\mu^u)\nonumber\\
& &\ \
+(-{\textstyle {1\over2}}+{\textstyle
{2\over 3}}\sin^2\theta_W)
      		(J_\mu^d+J_\mu^s),
		\eqnum{4.4c}
\end{eqnarray}
are the Standard Model electromagnetic and weak vector currents in
terms of quark field operators. (We ignore quarks heavier than
strange.) $G_E^s$ is thus the strangeness electric form factor, and is
constrained to be strictly 0 at $q^2=0$.  Note that one recent estimate
\cite{jaf} gives a strangeness mean-square-radius of around 0.14
fm$^2$, roughly as large as that for the neutron electric charge (but
of opposite sign). (This quantity can in principle be measured in,
e.g.  parity violating ${\vec e}^{\,-}$ scattering from nucleons at
forward angles.)

With the above relations, we see immediately that at $q^2=0$, the usual
weak charges are exactly obtained:
\begin{eqnarray}
Q_p^{w}\ =&&\ {\textstyle{1\over2}}(1-4\sin^2\theta_W),
\eqnum{4.5a}\\
Q_n^{w}\ =&&\ -{\textstyle{1\over 2}},	\eqnum{4.5b}
\end{eqnarray}
and one can also predict weak rms radii
\FL
\begin{eqnarray}
\left\langle r^2\right\rangle_{I,p}^{w}\ =&&\
{\textstyle{1\over2}}(1-4\sin^2\theta_W)
\left\langle r^2\right\rangle_{I,p}^{\gamma}
      			-{\textstyle{1\over 2}}
\left\langle r^2\right\rangle_{I,n}^{\gamma} \nonumber\\
 &&\qquad -\
	{\textstyle{1\over 2}}\left\langle r^2\right\rangle_I^{s}
	 +6({\textstyle{1\over2}})
(1-4\sin^2\theta_W)/(8M^2), \nonumber\\
&& \eqnum{4.6a}\\
\left\langle r^2\right\rangle_{I,n}^{w}\ =&&\
{\textstyle{1\over2}}(1-4\sin^2\theta_W)
\left\langle r^2\right\rangle_{I,n}^{\gamma}
      			-{\textstyle{1\over 2}}
\left\langle r^2\right\rangle_{I,p}^{\gamma} \nonumber\\
&&\qquad -\
	{\textstyle{1\over 2}}\left\langle r^2\right\rangle_I^{s}
	+6(-{\textstyle{1\over2}})/(8M^2). \eqnum{4.6b}
\end{eqnarray}
where the  subscript I indicates intrinsic nucleon structure, and the
last terms in (4.6a) and (4.6b) are the inclusion of the small
Darwin-Foldy correction to the radii.  Note that the neutron
(electromagnetic) contribution to the proton weak radius is not
suppressed by any $(1-4\sin^2\theta_W)$ factor, and thus is
surprisingly significant.

Using $\sin^2\theta_W\approx .23$,
$\left\langle r^2\right\rangle_{I,p}^{\gamma}\approx 0.7 $ fm$^2$,
$\left\langle r^2\right\rangle_{I,n}^{\gamma}\approx -0.11 $ fm$^2$,
$\left\langle r^2\right\rangle_I^{s} = 0$ gives
\begin{eqnarray}
\left\langle r^2\right\rangle_{I,p}^{w}\approx & &\
{\textstyle{1\over2}}(1-4\sin^2\theta_W)(2.1\ {\rm fm}^2), \eqnum{4.7a}\\
\left\langle r^2\right\rangle_{I,n}^{w}\approx & &\ -
{\textstyle{1\over2}}(.74\ {\rm fm}^2), \eqnum{4.7b}
\end{eqnarray}
The quantities in parentheses above can be interpreted as the physical
size (squared) of the weak distributions.  Note that using numbers of
${ \cal O}(\pm .1)$ \cite{jaf} for the strangeness radius will have
a large effect on $\left\langle r^2\right\rangle_I^{w}$ for both the
proton and (somewhat less so) the neutron.

To a good approximation, considering only rms radii, but no higher
moments, the relevant PNC matrix element is then given by a convolution
of (point) nucleon centers with their intrinsic structure, yielding a
replacement for Eqs.  (2.3) and (2.8),
\FL
\begin{eqnarray}
&&<\!{i}|H_{PNC,1}|{j}\!>\ \nonumber\\
&&\ \ ={G_F\over \sqrt2}\int
 <\!\Bigl(NQ_n^w\rho_n(\vec r\,)+
ZQ_p^w\rho_p(\vec r\,)\Bigr)
\psi_e^\dagger\gamma^5\psi_e\!> d^3\vec r\,,\nonumber\\
&&\eqnum{4.8a}\\
 &&\ \ ={G_F\over \sqrt2}C_{ij}(Z){\cal N}
      	[N Q_n^{w} q_n + Z Q_p^{w} q_p],\eqnum{4.8b}
\end{eqnarray}
with the quantities $q_p$ and $q_n$ slightly modified from eqns. (2.9),
\FL
\begin{eqnarray}
q_{(p,n)}=\int\,d^3\vec r\,&&\biggl(\rho_{(p,n)}^{c}(\vec r\,) \nonumber\\
&&\ \quad +\ {\textstyle{1\over 6}}\left\langle
 r^2\right\rangle_{I,(p,n)}^{w}\nabla^2\rho_{(p,n)}^{c}/
Q_{(p,n)}^{w}
      		\biggr)\, f(r),	\nonumber\\
&& \eqnum{4.9}
\end{eqnarray}
where $\rho_{p,n}^{c}(\vec r\,)$ is now the
density distribution of nucleon
{\it centers}, normalized to 1.

Assuming, for simplicity, uniform nucleon distributions, with
$R_p \approx R$,
\FL
\begin{eqnarray}
q_p\ 	\approx\ 1-(Z\alpha)^2\biggl(&&
      	.26\      +\ {.32\over R^2}
	\bigl(2.1 -\left\langle r^2\right\rangle_I^s/2Q_p^{w}\bigr)
      					 \biggr),\nonumber\\
q_n\ 	\approx 1-(Z\alpha)^2\biggl(&&
      	.038 +.221{R_n^2\over R^2}\nonumber\\
 	&&\qquad\ \ \ +\ {.32\over R^2}
      	\bigl(.74 - \left\langle r^2\right\rangle^s_I/2Q_n^{w}\bigr)
     	      				\biggr)\,,\nonumber\\
&& \eqnum{4.10}
\end{eqnarray}
with all radii measured in fm.

For $^{208}$Pb, the internal nucleon structure contributes about 0.002
to $q_n$, and a possible strangeness radius discussed above (0.14
fm$^2$) would contribute about 5 times less.  The internal structure
corrects $q_p$ by about 0.005, and the strangeness contribution here
would be comparable, about 0.004.  In Cs, these numbers turn out to be
smaller by about 40\%.

{}From the discussion to come in Sec. VI, we will see that these
contributions from (known) finite nucleon structure contributes at
about the 0.2\% level in an extraction of the weak nuclear charge when
measuring a single isotope of Pb (0.1\% level for Cs).  This might need
to be taken into account in an extremely high precision analysis, but
it will not add to the {\it uncertainty} in testing the Standard
Model.  (See also the complete discussion in Sec. VI to compare with
the expected scale of nuclear, atomic, and electroweak radiative
corrections and uncertainties.) On the other hand,  strangeness
contributions, which are currently very uncertain, might affect a
determination of the weak charge at below the 0.1\% level in Pb, and
even less in Cs, and thus are likely to be quite negligible.  They
could only become relevant if the nucleon strangeness radius were
comparable to the electromagnetic radius itself.

In the case of isotopic ratios, the internal nucleon structure plays an
even smaller role.  This is because errors then come from uncertainties
in the difference $q_n'-q_n$ (see Sec. VI).  To a good approximation,
nucleon structure effects are simply additive in mean square radii and
thus cancel in the differences.  Thus neither nucleon structure, nor
the uncertainties therein, are significant when extracting
$\sin^2\theta_W$ from isotope ratios.

\section{ NUCLEAR MODELING}

{}From the rather simplistic model of section III, we already observe
that the desired high precision measurements of electroweak parameters  will
require knowledge of neutron radii in heavy nuclei to within at least
several percent. (see also the discussion in section VI) At this level,
one clearly must treat higher moments with some care, and the
microscopic details of the nucleon distributions may be of some
importance.  For this reason, we have attempted to evaluate $q_n$ and
$q_p$ numerically, utilizing the best existing nuclear models for
neutron, proton, and charge distributions available to us.  In this
section, we discuss some of the basic features of these models, along
with some caveats on their reliability for neutron observables.

The nuclear many-body problem presents a formidable challenge for
infinite nuclear matter, and an even greater one for heavy finite
nuclei.  The most popular  route being taken today is some version of
Hartree-Fock, which has had considerable success in describing a
variety of nuclear properties semiquantitatively.

\subsection{ Brueckner-Hartree-Fock}
The underlying basis of nuclear Hartree-Fock calculations is Brueckner
theory.  The elementary two-body interactions are too strong
(especially the short-range repulsion) to lead to meaningful HF
calculations.  Although there has been extensive work on nuclear matter
calculations using Brueckner theory and beyond, for finite nuclei only
light ones have been considered \cite{kb} and nothing for the
nuclei of interest here.

The lowest level is the independent pair approximation.  The {\it
effective} interaction is not $v$ but the Brueckner $G$-matrix, where
$G=v\,F$, with $\psi(1,2)=F(1,2)\phi(1)\phi(2)$.  $G$ satisfies a
scattering-type equation with a projection operator in intermediate
states which excludes scattering back into the Fermi sea; this also
results in no phase shift (for pairs in occupied states) and to the
two-body wave function ``wound" which extends over a``healing
distance." The $\phi$'s are to be identified with the HF single {\it
quasi}-particle functions.  We have used $F$ here to denote the
two-particle correlation function.  For repulsive core potentials, $F$
has a hole (wound) centered about $r=0$.  The Brueckner $G$ is
non-local, and both energy- and density-dependent.

\subsection{ Two-body correlations}
The dependence of the single
particle density distribution on the correlation
function is relatively small.
We can estimate it as follows.  Let
\begin{equation}
\rho_2(\vec r_1,\,\vec r_2)=\rho_1 (r_1)\rho_1
(r_2)f(\vec r_1-\vec r_2)\,.\eqnum{5.1}
\end{equation}
Let $\rho_1(r)\propto e^{-r^2/a_N^2}$ and $f(r)\propto e^{-r^2/a_c^2}$
For $a_c<<a_N$, one finds for the rms size of the single particle density
distribution
\begin{eqnarray}
<\! r^2 \!>\ &=&\  \int d^3r_1\,d^3r_2\,r_1^2\,
\rho(\vec r_1,\vec r_2)\nonumber\\
&\approx&\  <\! r^2 \!>_1\left[1+{1\over
4\sqrt{2}}
\left({a_c\over a_N}\right)^3\right]\,,\eqnum{5.2}
\end{eqnarray}
where $<\! r^2\!>_1$ corresponds to $\rho_1$.
For heavy nuclei (say $a_N / a_c\approx 7.0/0.7$ fm), the
correction is less than $2\times 10^{-4}$, which is below our level of
current concern.

\subsection{Phenomenological Hartree-Fock, including
deformation and pairing}
Because of the numerical complexity, most HF calculations have
employed phenomenological potentials intended to simulate the Brueckner
$G$-matrix.  The most commonly used potentials are varieties of the
very convenient Skyrme interaction.  The Skyrme interactions are of the
delta-function form and as such lead to single particle equations with
local one-body potentials and spatially-dependent effective masses,
with no more complication than Hartree calculations.  In contrast,
finite range interactions lead to non-local single particle potentials
arising from the exchange term.  Momentum-dependent Skyrme interactions
do not lead to further complications and simulate some effects of
finite range.  Calculations have also been done with finite range
forces, using e.g. the Gogny interaction\cite{dec} Note that none
of these phenomenological potentials are intended to reproduce free
nucleon-nucleon scattering.  There are of the order of eight (more or
less) adjustable parameters in any model.\cite{que}

      Most nuclear structure calculations on heavy nuclei are carried out
in the deformed Hartree-Fock or the Hartree-Fock-Bogolyubov
approximations.  The latter include BCS-type pairing.
 Hartree-Fock encompasses a limited class of correlation structure.  Only
correlations of a collective nature are included.  It is not surprising
to find that in HF neutron and proton densities tend to track one another.
Nevertheless, they do exhibit the expected behavior that the neutron rms
radius increases more rapidly than the proton one with increasing $A$ in an
isotopic sequence.  The relative tracking (variation in the neutron
skin) depends on the way
in which symmetry energy is handled.

Intrinsic deformations play a key role in spherically averaged
proton and neutron densities.  In the uniform, incompressible
approximation, for example, the mean square radius is increased
according to\cite{whf}
\begin{equation}
<\! r^2\!>_\beta=<\! r^2\!>_0 \left[1+{5\over 4\pi}
<\! \beta^2\!>\right] \eqnum{5.3}
\end{equation}
where $\beta$ is the nuclear shape parameter, proportional to the
quadrupole moment.  $\beta$ can attain values of the order of $1/3$ and
changes in $<\! \beta^2\!>$ among isotopes can produce deviations in
spectroscopic isotope shifts by an order of magnitude from the
$A^{1/3}$ law.  Although HF calculations tend to yield spherical
($<\!\beta=0\!>$) or near spherical equilibrium shapes for the Pb
isotopes, $<\!\beta^2\!>$ is not zero and changes in $<\!\beta^2\!>$
have been considered by some authors.

\subsection{Relativistic Mean Field}
Although nuclear structure is primarily non\-rel\-a\-tiv\-is\-tic,
con\-sid\-er\-able suc\-cess has been achieved by treating the nucleons
and protons as point Dirac particles\cite{ser} [see, however,
Achtzehnter and Wilets \cite{ach}] interacting with phenomenological
vector and scalar mesons in the mean field approximation.  The vector
mesons can be identified with the isoscalar omega and the isovector rho
mesons; the scalar meson is a simulation of two-pion  exchange.  The
mesons are treated in the mean field, or c-number approximation.  An
attractive feature of the model is that the strong spin-orbit potential
appears to emerge ``naturally."  While this turns out to be true for
isoscalar potentials, it fails badly for the isovector potentials, but
can always be parameterized to yield reasonable results.\cite{ach}

      In order to fit nuclear properties, it has been necessary to go
beyond linear field theory.  Self-interaction of the scalar field has
been introduced, with additional parameters.  Among other problems,
this solved the compression modulus anomaly, which is much too large in
the linear model. The total number of adjustable parameters which must
be introduced is comparable to that required in models using Skyrme
forces. As with Skyrme forces, the mean field approximation does not
lead to nonlocality in the one-body potentials.

\subsection{ RPA and MCHF}
The tail of the neutron or proton distribution has
correlation/polarization corrections not described by HF, at least for
large distances.  The reason for this is that the individual nucleon
wave functions see the potential of the ``mean" self-consistent core.
In the tail region, the residual core tends to relax.  This is most
evident for the separation energy:  In HF, the separation energy of a
nucleon is just the energy eigenvalue (Koopmanns' theorem) if the core
is frozen.  If the energy of the residual nucleus is recalculated self
consistently, the separation energy is reduced by what is termed the
rearrangement energy.  Correlations of this type are included through
RPA, which is equivalent to small amplitude, time-dependent
Hartree-Fock.

Other types of correlations could be included through
multiconfigurational Hartree-Fock, which, as the name implies, means that
the trial wave function is not a single determinant, but a sum of
determinants (configurations).  This serves two purposes:  correlations of
the kind allowed by the choice of configurations are included, and the
occupation of these configurations modifies the mean field potential and
single particle functions.

\subsection{Beyond Hartree-Fock plus}
Most HF and HFB calculations do reasonably well in reproducing
atomic isotopic shifts for the {\it even-even} isotopes of PB below
208.  So, incidentally, does the droplet model of
Meyers\cite{mey}.  They all fail to reproduce even-odd staggering,
which shows odd nuclei to be smaller than the mean of their even-even
neighbors, and also do badly on the shifts above 208.

There are no giant shell model diagonalization calculations available
which yield densities for heavy nuclei.  Such would be very valuable
for comparison with Hartree-Fock results, since in principle they
include all types of correlations, limited only by the size of the
basis.

An idealized shell model calculation should be based on realistic two
body interactions, the kind which fit free two-body scattering data.
The Hilbert space could be divided into a ``near" and a ``far" space.
The effective two body interaction could be obtained by solving for the
Brueckner $G$-matrix with the intermediate states excluded from the
near space.  The far space scattering states could be approximated by
plane waves if the momentum sphere separating the spaces is
sufficiently large.\cite{kb} The Hamiltonian matrix for the
inner space, using the effective interaction, is then diagonalized.

\subsection{Summary and Discussion}
Unfortunately, not all of the theoretical considerations discussed
above have been incorporated in any single calculation.  Heavy nuclei
pose a difficult challenge for reliable, detailed modeling at the level
of precision we require.  There do exist in the literature a number of
recent efforts, as discussed in sections C and D above, which involve
either relativistic or non relativistic Hartree-Fock nuclear calculations.
We have accumulated densities from several of these authors in order to
evaluate $q_p$ and $q_n$ and make comparisons among the different
models.  These include various HF calculations with Skyrme
forces,\cite{gar,fri} an HFB calculation with a Gogny finite
range D1S interaction,\cite{dec,gir} and several relativistic
mean field models.\cite{rei,reiII} The results are presented
in Sec. VII.

  The modelers fit their adjustable parameters to choices among various
bulk properties (energy per nucleon, compressibility modulus, symmetry
energy, etc.), and properties of particular nuclei (energies, charge
radii, deformations, spectra, multipole sum rules, etc.)  Indeed, the
physics behind the models comes, in part, from the choice of the
particular observables included in the parameter fits.  The models we
have selected all do roughly equally well in fitting the wide range of
nuclear observables available across the periodic
table.\cite{fri,rei}

The analysis of atomic PNC, as discussed in sections II and III, relies
on a detailed knowledge of neutron distributions in nuclei. The lack of
unambiguous, precise experimental measures of neutron radii means that
all of these models must ``extrapolate'' to the desired neutron
properties.  Charge radii, on the other hand, are in a certain sense
``built in'', in that the set of observables to which the nuclear model
parameters are fit includes charge radii of several even-even nuclei,
one of which is $^{208}$Pb.  In defense of the models, they predict
with good success the charge radii of other even-even nuclei not
included in the fit,\cite{reiII} and also reproduce the
well-measured isotopic charge radius shifts for, e.g. the even isotopes
of Pb.\cite{gar,auf} However, they do not reproduce the
observed even-odd staggering of charge radii very well, nor are the
results as good for the charge radius of $^{210}$Pb, an indication that
some care should be taken when considering non closed-shell cases.

There do exist some data which may give more direct information on the
neutron skin. This might be used as additional input to these nuclear
models, and could help further constrain the predictions for neutron
radius, and neutron isotope shifts, if one could demonstrate
consistency in the results.  Perhaps the best known data comes from 800
MeV polarized proton scattering from
$^{208}$Pb.\cite{ray,batty} This gives $R_n - R_p = 0.14 \pm
0.04$ fm. The quoted error, which is quite small for our purposes,
contains both statistical and certain theoretical uncertainties  as
stated.  However, there are still {\it additional} theoretical
uncertainties, involving e.g. assumptions about the {\it in medium}
nucleon-nucleon t-matrix, and the result seems to exhibit a rather
large and troubling energy dependence.\cite{rayii}  The
absolute value of $R_n$ in one isotope is believed to  be fairly
difficult to obtain with confidence using such experiments. However, it
may be that the relative shift in $R_n$ among isotopes involves
cancellations that reduce these theoretical uncertainties. Measurements
on other Pb isotopes (data\cite{hin} apparently exists for $^{206}$Pb)
would clearly be of interest in this context.  Further
experimentation and theoretical analysis at other energies are also
crucial to demonstrating the consistency of the results.

Experiments involving intermediate energy charged pion scattering from
nuclei may also help further constrain the neutron radii, or
the relevant isovector model parameters, as well. Such data exists for
$^{208}$Pb,\cite{olm} but again the absolute normalization
poses a real challenge to analyses.\cite{batty} Taken at face
value, the $\vec p$ and $\pi$ results for lead neutron radii do agree
with one another reasonably well, and also match with e.g. the Gogny
finite range Hartree-Fock calculations. Another experimental
possibility involves energies and sum rule strengths of giant multipole
resonances.\cite{kra} The uncertainties here are even larger,
and difficult to estimate.  Clearly, a reliable set of such additional
``strong probe'' inputs, including yet other options such as $\alpha$
particle scattering, kaon scattering, Coulomb displacement energies,
etc., could be an aid in constraining the theoretical models
on neutron properties.

Another promising experimental possibility for the future might be
direct electroweak experiments, such as parity violating asymmetries in
elastic, intermediate energy  ${\vec e}^{\,-}$ - nucleus scattering, as
proposed by Donnelly, Dubach, and Sick,\cite{donn} or perhaps
elastic $\nu$ scattering. The reactions and analyses are quite clean,
just as in the charge scattering case. There would be, for example, no
serious ambiguities in the absolute scale of the radii measured.  Such
experiments would in fact be sensitive to the full nuclear weak charge
distribution,  rather than just the RMS radius, which many of the
strong probe measurements are primarily sensitive to.  Because such
experiments could be done at moderate momentum transfers ($q \sim 1 $
fm$^{-1})$, the extraction of nuclear distribution information would be
much less sensitive to the precise values of electroweak parameters
than in the corresponding atomic parity violation case.  The
asymmetries and $\nu$ cross sections are naturally  extremely small,
and the experimental challenges are formidable. Nevertheless, recent
estimates for the parity violating asymmetries indicate that
measurements sensitive e.g.  to the neutron RMS radius in $^{208}$Pb at
the 1\% level are feasible.\cite{donn} As we will see in
sections VI and VII, such a level would make the nuclear structure
uncertainties quite negligible for the purposes of extracting standard
model parameters from single isotope atomic PNC measurements.

In any case, current model fitting has been done  with the best and
most reliable data at hand, most of which are  not directly sensitive to
neutron distributions. It is always difficult to estimate the
theoretical uncertainties in such model calculations.  In this
section, we have already mentioned several potentially important
missing features that future work should address, especially involving
 nucleon correlations.  We have not attempted
here to try to choose a ``best model'' from the various ones we
examined, but rather wish to evaluate the existing spread in
predictions as an effective lower bound on the theoretical uncertainties
involved.  One might, however, try to make a selection based on
detailed comparisons, specifically targeting a good fit to
heavier nuclei energies, isotopic shifts, giant dipole properties, and
other quantities potentially sensitive to isovector properties.  We
encourage work in such directions. The goal should be to find the most
reliable model(s) while still retaining an estimate of the remaining
theoretical uncertainties.

\section{ERROR ANALYSIS AND TESTS OF ELECTROWEAK PHYSICS}

One of the motivations for further improving atomic parity violation
experiments is to test the Standard Model at the level of its
one-loop electroweak radiative corrections. This allows one to probe
for possible small ``new physics'' effects, which would appear
as further loop corrections or more directly as additional
interactions at the tree level (i. e. without loop corrections).  A
good example of the latter, to which atomic PNC is particularly
sensitive, is exchange of a second, more massive, neutral Z-boson
required in theories with larger gauge groups. In the Standard
Model, the loop contributions are separated into two parts: fixed
radiative corrections due to contributions from the known quarks, leptons
and bosons, and the heavy physics part due to contributions from the top
quark and the Higgs boson.  One is interested in an experimental
determination of the heavy physics part, which in the language of
Marciano and Rosner\cite{mr} is expressed in terms of weak
isospin-conserving, S, and isospin-breaking, T, effects.  These two
constants\cite{pes} are a convenient way not only of including
uncertainties in the top quark and Higgs masses, but of
parameterizing the effects of some specific classes of new physics
as well.  It turns out that low energy PNC measurements are nicely
complementary to high energy measurements such as direct Z-boson
production, since both the radiative corrections and the
sensitivities to new tree-level interactions are quite different.

To show how the radiative corrections and the possible new physics
enter into atomic PNC, and how they might compare in size to nuclear
structure effects, we begin by rewriting equation (2.9) in the form:

\FL
\begin{equation}
<\! i|H_{PNC,1}|j\!>={G_F\over 2\sqrt{2}}
C_{ij}(Z){\cal N}
   [Q_W(N,Z)+Q^{nuc}_W(N,Z)] \eqnum{6.1}
\end{equation}
where $Q_W(N,Z)$, known as the nuclear weak charge, is the quantity of
primary interest to electroweak theory, and in the standard model
without radiative corrections reduces to:
\begin{equation}
Q_{W}^0=-N+Z(1-4\,\bar x)  \eqnum{6.2}
\end{equation}
where $\bar x \equiv\sin^2\theta_W$. $Q_W(N,Z)$ is determined from
atomic experiments by combining atomic measurements of $<\!
i|H_{PNC,1}|j\!>$ with calculations of both atomic structure (contained
in the factor $C_{ij}$) and nuclear structure.  The nuclear structure
corrections are contained in $Q^{nuc}_W(N,Z)$, which is given by:
\FL
\begin{eqnarray}
Q^{nuc}_W(N,Z)\ &\equiv&\ Q_W([q_n-1]N,[q_p-1]Z)\nonumber\\
&\approx&\  -N(q_n-1)+Z(1-4\,\bar x)(q_p-1)\,. \eqnum{6.3}
\end{eqnarray}
Nuclear structure is also contained in the normalization $\cal N$, but
as we will see in section VII, $\cal N$ is determined by the nuclear
charge distribution, which is usually known experimentally.

When we include possible new physics, together with the effects of
radiative corrections which have been calculated by
others,\cite{mr} $Q_W(N,Z)$ becomes:
\FL
\begin{eqnarray}
Q_W(N,Z)=&&(0.9857\pm0.0004)(1+0.00782T)\nonumber\\
&&\ \times\ [-N+Z(1-(4.012\pm0.010)\bar x] \nonumber\\
&&\ \ \ \qquad + Q^{new}_{tree}(N,Z)   \eqnum{6.4}
\end{eqnarray}
\narrowtext
where $\bar x$ is assumed here to be defined at the mass scale $m_Z$
by modified minimal subtraction,\cite{mr} and is given by:
\begin{equation}
\bar x =
.2323 \pm .0007 +.00365S-.00261T\,.  \eqnum{6.5}
\end{equation}
The errors indicated in (6.4) and (6.5) come from uncertainties both in
experimental input parameters and in evaluations of known physics
loop-diagrams.  The unknown, heavy physics loop-corrections are
contained in the parameters $S$ and $T$, which depend upon the heavy
masses, and are defined such that $S = T = 0$ if $m_H =100$ GeV, $m_t
= 140$ GeV, and if there is no new physics beyond the Standard
Model.  Including $Q^{new}_{tree}(N,Z)$ in $Q_W(N,Z)$ allows for additional
tree-level physics beyond the Standard Model.  For example, exchange
of the extra $Z_x$ in SO(10) models\cite{mr,lon} (assuming no $Z_x-Z$
mixing) would make:
\begin{equation}
Q^{new}_{tree}(N,Z)\approx
0.4(2N+Z)m^2_W/m^2_{Z_\chi}\,.    \eqnum{6.6}
\end{equation}

It is useful to consider how well the parameters in $Q_W$ are currently
known.  The central value of $Q^{new}_{tree}(N,Z)$, determined mainly
by Cs PNC measurements, is about $2.2 \pm 1.6 \pm .9$ ({\it if} all
other heavy physics in equation (6.4) is ignored), and corresponds in
the SO(10) model to $m_{Z_\chi}\approx 500$ GeV. Conversely, assuming
no new tree-level physics (i.e., $Q^{new}_{tree} = 0$), the experimental
uncertainty in T  is currently around $\pm 1$, and in S around $\pm 3$,
the latter determined largely from Cs PNC. Ultimately, as Marciano and
Rosner have indicated, an effort to reduce the uncertainty in S to $\pm
0.2$ is extremely important, since at that level it is sensitive even
to minimal one-doublet technicolor models. This sort of accuracy is an
extreme challenge to either high energy or atomic experiments.  Current
knowledge of $\sin^2\theta_W$ from a global analysis of electroweak
data\cite{ken}can be summarized by $\bar x = 0.230 \pm .004$,
(roughly 2\% uncertainty).  If future high energy measurements  were to
reduce the uncertainty in $\bar x$ beyond what is attainable in atoms,
the atomic experiments would still be valuable for improving the limits,
e.g., on an additional Z.

In summary, any improvement in determining atomic PNC is likely to
provide useful information about electroweak physics, and it becomes
extremely important to work out how much nuclear structure
uncertainties may be a limiting factor, and to reduce these
uncertainties where possible.

We first consider the impact of nuclear uncertainties on
PNC measurements of single isotopes.  PNC experiments to date have
been done on stable isotopes of heavy atoms, namely Cs, Pb, Bi, and
Tl, and have not compared different isotopes of the same element.
{}From equation (6.1) we derive an expression for the uncertainty in
$Q_W$ in terms of the uncertainties in atomic and nuclear structure
and in the measured quantity ${\cal O} \equiv <\!i|H_{PNC,1}|j\!>$:
\begin{equation}
{
{\delta Q_W \over Q_W} \ \approx\
{\delta \cal O \over \cal O} - {\delta C_{ij} \over C_{ij}} -
{\delta \cal N \over
\cal N} - {\delta Q^{nuc}_W \over Q_W}  \,.    \eqnum{6.7}
}
\end{equation}

 If we assume that $\cal O$ can be measured to arbitrary
accuracy, and that proton distributions (which will
influence $\delta {\cal N}$) are also well enough understood and/or
measured, there remain the uncertainties coming from atomic and
nuclear structure, which we can write in the form:
\begin{equation}{
{\delta Q_W \over Q_W} \ \approx\  - {\delta C_{ij}
\over C_{ij}} - {\delta q_n}    \eqnum{6.8} }
\end{equation}
where we have dropped all terms containing the factor $1-4.012\bar
x$, which should be quite negligible due to the accidental value of
$\bar x \approx {1\over 4}$.  Rewriting in terms of the weak
interaction parameters, we obtain:
\begin{equation}
0.014{Z\over N}\delta S +
{\delta Q^{new}_{tree}(N,Z)\over Q_W}
\approx\  -{\delta C_{ij} \over C_{ij}} -
{\delta q_n} \,.    \eqnum{6.9}
\end{equation}
ignoring the contribution of the weak-isospin breaking parameter T
which cancels to better than 10\% for the full range of $Z/N$
found in the elements of experimental interest.  Thus a PNC
measurement in a single isotope can set limits on the
weak-isospin conserving parameter $S$ and/or new tree-level
interactions, and in fact the best limits on both of these
parameters now come from PNC measurements in atomic cesium.  To
determine the role of nuclear structure, we must compare the
uncertainty $\delta q_n$ on the right hand side of equation (6.9) to
the atomic structure uncertainty $\delta C_{ij}/C_{ij}$ .  This we do
later, in section VII.

Because of the difficult atomic physics calculations, there has been
some serious interest in measuring parity violation in a chain of
isotopes. Taking ratios between isotopes cancels essentially all
dependence on atomic structure. Unfortunately, although the
atomic physics indeed cancels in the ratio, the nuclear structure does
not.  Referring to equation (6.1) we consider the ratio:
\begin{equation}
{\cal R} \equiv {{\cal O}\over {\cal O'}}
 = {{[Q_W(N,Z)+Q^{nuc}_W(N,Z)]{\cal N}}\over
{[Q_W(N',Z)+Q^{nuc}_W(N',Z)]{\cal N'}}}\,.   \eqnum{6.10}
\end{equation}
where primed and unprimed quantities refer to different isotopes.
The sensitivity of $\bar x$ and $Q^{new}_{tree}(N,Z)$, extracted from
this ratio, to the nuclear structure is then given approximately by
\begin{eqnarray}
Z{\delta \bar x\over \bar x}\ &&- \delta
Q^{new}_{tree}(N,Z) + N{\delta \Delta Q^{new}_{tree} \over \Delta N}
\nonumber\\
&& \approx\ {NN'\over \Delta N}
   \left[  {\delta {\cal R}\over {\cal R}}
        + {\delta (\Delta {\cal N})\over {\cal N}}
      + {\delta (\Delta q_n)      \over q_n} \right]. \eqnum{6.11}
\end{eqnarray}
\narrowtext
where we have made simplifying assumptions that the isotopes are close
together, i.e. $\Delta N \equiv N'-N << N$, that $\sin^2\theta_W
\approx 1/4$, and where we have used e.g.
$({\delta q_n/ q_n}-{\delta q_n'/q_n'})\approx{\delta(\Delta q_n)/q_n}$,
which is numerically accurate for the models of Pb we have considered.
Because of the special sensitivity of atomic PNC to any additional
heavy Z-bosons, we note as an example that a determination of $m_{Z_x}$
in the model of equation (6.6) would be constrained by replacing the
left hand side of equation (6.11) by
$ \left( Z{\delta \bar x / \bar x}\ - 0.4Z \delta
{({m^2_W / {m^2_{Z_x}}})} \right)	$.

The uncertainties on the right side of (6.11) are
effectively in the relative {\it difference} between
quantities for two isotopes. In principle, different nuclear models
which disagree on the absolute values of, say, $q_n$ may agree on the
relative {\it change} in this quantity to a much higher degree of
accuracy.  However, such a reduction in uncertainty in the terms
within the  brackets in expression (6.11) is roughly compensated
by the factor $N/\Delta A$.
Comparing with equation (6.9) for a single isotope, in which any
new tree-level interactions enter with the equally uncertain
loop-parameter $S$, we see that when we instead compare isotopes,
$Q^{new}_{tree}$ appears together with a different parameter, $\bar
x$, which is independently measurable in high energy experiments.

\section{DISCUSSION}

To calculate $q_n$, we use various theoretical predictions of neutron
and proton distributions from the literature.  Proton distributions are
used to compute   $f(r)$, the electronic wave function overlap
defined in equation (2.5).  We solve numerically  for single electron
Dirac $s_{1/2} $ and $p_{1/2}$ wave functions near the origin, in the Coulomb
potential of the nuclear charge distribution (as discussed in section
II), and make {\it no} approximation of a power series in $Z\alpha$, as
was done e.g. for equation~(3.2).  We have neglected the contributions
to the nuclear charge distribution from internal neutron structure,
as discussed in section IV.  We estimate the error associated with
this assumption to be well below the level
of the model uncertainties themselves.  The quantity $q_n \equiv
\int\,\rho_n(r)f(r)\,d^3r$ is then calculated directly from the
corresponding neutron distribution.

In Tables \ref{table1a} and \ref{table1b},
we present the rms radii $R_n$ and $R_p$, the correction
factors $q_n$ and $q_p$, and the electron normalization $\cal N$, for
several nuclear models of the Pb isotopes 202 and 210.
Except for the norm, the  spread in values in the final rows
of Tables \ref{table1a} and \ref{table1b}
should give some indication of a {\it lower limit} on the current
level of theoretical model-dependent uncertainties, assuming that one
accepts these models as equally phenomenologically reasonable.

The normalization factor $\cal N$ defined in equation~(2.5), which is
proportional to $1/f(r\gg R_p)$, is defined arbitrarily here as
$0.10361/f(300 \ \rm fm)$. The numerator, $f^{\rm\,expt}(300 \ \rm
fm)$, is evaluated using a model independent experimental charge
distribution from electron scattering off $^{208}$Pb. As stated
earlier, we are not concerned with the absolute value of the norm, but
only its dependence on atomic weight and charge distribution. This
definition simply scales $\cal N$ to be near 1.0.  For $^{208}$Pb, the
model spread in the normalization from Table \ref{table1b} might appear
to contribute at a significant level.  One can, however, consider
correcting $\cal N$ by using an approximate formula relating $\cal N$
to the charge radius, namely ${\cal N} = R^{-\gamma}$, as in
eqn.~(2.6).  This is given by
\begin{equation}
{\cal N'} = {\cal N}
(R^\gamma/R^\gamma_{\rm expt}).   \eqnum{7.1}
\end{equation}
The model spread in
this ${\cal N'}$ is significantly reduced.  The point is that these
models are not precisely reproducing the observed charge radii of the
lead isotopes, which feeds rather directly into a calculation of $\cal
N$.  The correction factor above compensates for this, using the
existing high precision measurements of charge radii from optical
isotope shifts and electron scattering.\cite{auf,jag}

Some of the results in tables \ref{table1a} and \ref{table1b}
are reproduced in graphical form also, in Figures \ref{fig1a},
\ref{fig1b}, and \ref{fig2}.  In Fig.~\ref{fig1a}, we plot the
predicted $R_n$ versus atomic weight for several even lead isotopes,
and in Fig.~\ref{fig1b} the ratio $R_n/R_p$.  The spread in $R_n$ among
models is decidedly larger than the spread in $R_p$.  We do note a
systematically larger neutron radius in the relativistic
models.\cite{sha}  The origin of this is indeed not yet completely
understood, but may be connected with larger asymmetry energies found
in these models. This in turn might tend to pull neutron and proton
distributions together where the densities are high, leaving a somewhat
larger neutron tail.

In Fig.~\ref{fig2}, we plot $q_n$ versus atomic weight for several even lead
isotopes. The spread is closely related to the spread in $R_n/R_p$
shown in Fig.~\ref{fig1b}, as might be expected from the simplified
formulas (1.1) or (3.3b) based on uniform nuclear charge density.
Estimates of $q_n$ using these simplified formulas yield the same
general trends as the detailed calculations, with absolute values
differing generally by parts in a thousand or less. The relativistic
models yield somewhat smaller $q_n$, due to their larger $R_n/R_p$
ratio.

In the case of {\it single isotopes}, the total nuclear model spread
does not appear to be the most serious problem in using equation (6.8)
or (6.9) to extract weak interaction parameters from atomic PNC .  For
$^{208}$Pb,  the typical full spread in calculated $q_n$ is
$\alt 0.005$.  In the case of $^{133}$Cs, the sensitivity to nuclear
structure is even weaker, due to the smaller value of (Z$\alpha$). A
larger uncertainty, at least at the present time, is due to atomic
physics calculations.\cite{bjs,dfs} For example, Cs is one
of the most favorable elements from the point of view of atomic
theory, and to achieve the current level of quoted uncertainty of
$\delta C_{ij}(Z)/C_{ij}(Z) \approx $1\% in Cs is an impressive
task.  But this uncertainty is still probably larger than the
uncertainty in $q_n$ for Cs.  Significant future improvement in atomic
calculations is likely to be difficult. Thus, aside from any
experimental uncertainties, atomic structure is the present limiting
factor in getting $Q_W$ and the associated weak parameters from single
isotope atomic PNC measurements, and appears to remain so even after
considering the possible nuclear physics effects.  This conclusion
is consistent with the findings mentioned in the calculation of
reference 2.

Consider next the ratios in an isotopic chain, for example
($^{202}$Pb/$^{208}$Pb).  Referring to equation (6.11) we see that the
PNC experiments would then be measuring $\bar x$ and/or observing new
tree level physics.  For definiteness let us assume no new tree level
physics.  Then a $\pm$1\% extraction of $\bar x$ would require $\delta
\Delta q_n \alt 6\cdot 10^{-4}$, and ${\delta \Delta {\cal N}/ {\cal N}
} \alt 6\cdot 10^{-4}$.  Assuming uniform nuclear distributions, this
implies $\delta\Delta(R_n/R_p) \alt 4\cdot 10^{-3}$.  Referring to
Table \ref{table2}, which shows the change in various quantities
between these two particular lead isotopes, the model spread for
$\delta \Delta q_n$ is around $9\cdot 10^{-4}$ and  for $\delta \Delta
{\cal N}/{\cal N} $ is about $6\cdot 10^{-4}$.  Note however that when
${\cal N}$ is corrected as in (7.1) above, using experimental knowledge
of charge radii, this spread, at least, is significantly reduced to
below the $\pm$1\% level. This is seen from the final column in Table
\ref{table2}.  But the spread in $\Delta q_n$ remains, and is
comparable to the accuracy needed for a 1\% extraction of $\bar x$.
Similarly, the model spread in $(R_n/R_p)$ from Table III is about
5$\cdot10^{-3}$, which likewise corresponds to a $>$1\% spread in $\bar
x$.

If we exclude the relativistic models, which seem to have
substantially different neutron radii from the conventional H-F
calculations, the model spread just among the various Skyrme
parameterizations considered gives $\delta \Delta q_n \approx
4\cdot10^{-4}$.  It thus appears unlikely that PNC measurements
comparing Pb isotopes could yield much better than  a 1\%
determination of $\bar x$, unless there is significant improvement in
understanding of nuclear structure.

The same results can be seen perhaps more clearly in Fig.~\ref{fig3},
which displays in graphical form the values of  $\Delta
q_n(202\rightarrow 208)$ from Table \ref{table2} versus the different
models considered. The spread in predictions of this quantity is
actually larger than 100\%.  Also shown in the figure is a typical
scale of 1\% in the weak angle. As noted already, the model spread is
too large for extractions of $\bar x$ at the sub 1\% level if one
cannot otherwise eliminate or improve any of the models used.  On the
other hand, the nuclear structure uncertainties may not preclude a
significant improvement in sensitivity to new Z bosons or other new
tree-level physics in equations (6.11), particularly if $\bar x$ is
determined well by high energy experiments.

Although the nonrelativistic models do appear to cluster together
somewhat, one should perhaps be a bit wary of their apparent
self-consistency.  For example, a modification of the coefficient of
the isovector (n-p asymmetry) surface term, a
$(\rho_p-\rho_n)\nabla^2(\rho_p-\rho_n)$ term in the Skyrme
Lagrangian,\cite{fri} has little effect on most bulk properties, and
hence on the goodness of the Skyrme fits.\cite{rei} This term, however,
does modify the neutron skin significantly.  Reinhard's rough
estimates  show that an uncertainty of $\Delta R_n\approx \pm .15$ fm
is not unreasonable.\cite{rei} This in turn can modify the quantity
shown in Fig.~\ref{fig3} by amounts of ${\cal O}(6\cdot 10^{-4})$,
larger than the spread in the given Skyrme models.  The relativistic
models do not have such flexibility, as the isovector rho couplings are
largely constrained by isotopic trends in ground state energies and
charge radii, but this is of course no guarantee that these models
correctly describe all isovector properties equally well.

In the case of Cs isotopes, accurate calculations for neutron radii (or
even proton radii) are difficult.  They have odd Z, and require
additional approximations to deal with unfilled shells, as well as
deformations.  The lack of success in predicting the even-odd
staggering of $\delta \langle r^2\rangle_{ch}$ in lead isotopes
indicates the seriousness of these problems. An estimate of the scales
involved, however, can be made using calculations with existing nuclear
codes.  One such result\cite{bei} gives $R_n/R_p\approx 1.03$ for
$^{135}$Cs, and $\Delta R_n/R_p$ ($^{131}$Cs $\rightarrow \ ^{135}$Cs)
$\approx 5\cdot 10^{-3}$. If this latter number itself has a 100\%
uncertainty (for comparison, see Fig.~\ref{fig3} for the case of lead
which does show a 100\% spread among model predictions of the
equivalent quantity $\Delta q_n$ for about the same $\Delta A/A$), then
the uncertainty in $\bar x$ from this fairly small range of isotopes
would be approximately 1\%.   $\Delta A$ of up to 10 or higher may be
experimentally possible for Cs, which might help to reduce the nuclear
physics uncertainties.  From the experimental side, the absence of
stable isotopic partners to $^{133}$Cs makes it difficult to obtain
values of $\Delta R_n/R_p$ from parity violating electron scattering,
or $\vec p$ elastic, or pion experiments, as may be possible for the
lead isotopes. Further work on theoretical estimates for Cs isotopic
radii is  clearly called for.

Given a set of experimental results for isotopic PNC ratios, one can
also consider a bootstrap procedure:  from atomic experiments over {\it
several} isotope differences, use the various models to extract the
weak mixing angle.  Then, only those models which yield the same
$\sin^2\theta_W$ for the various isotopic pairs are acceptable.
Unfortunately, the various nuclear models we have considered (for lead
isotopes near $^{208}Pb$) yield predictions for the PNC ratios which
are fairly linear with $\Delta A$. Since this prediction is also
roughly linear with $\sin^2\theta_W$, it appears that the various
nuclear models could be internally consistent, each yielding a unique
$\sin^2\theta_W$ but differing from model to model about the extracted
value. Of course, one cannot draw any firm conclusions about this until
after the data are known. There are indeed some slight deviations from
linearity, especially for non-closed shell isotopes, and one may be
able to take advantage of this. In essence, this bootstrap idea uses
PNC atomic isotope ratios themselves as our desired additional
constraint on neutron properties - with a
large enough set of PNC data, one could hope to simultaneously constrain
the nuclear model parameters {\it and} measure the weak mixing
angle.

\section{CONCLUSIONS}

 For the case of Pb, in order to extract
electroweak parameters from atomic PNC experiments  at a level of precision
which would be considered ``significant'' for testing the Standard
Model, we have shown that it is necessary to have confidence in the
isotopic relative neutron/proton radius shift, $\Delta(R_n/R_p)/\Delta
A$, to better than a few times $ 10^{-4}$. We have examined various
nuclear model calculations, and find that the spread in theoretical values
corresponds to an uncertainty in the weak mixing angle greater than
1\%, with the assumption that no new physics is present.
Without some further basis for discriminating among the various
models, the spread represents a lower bound to the uncertainties in
the calculated values.

The basic problem is essentially that the models have been
parameterized to fit properties like charge distributions, which are
not directly sensitive to neutron distributions.  As Reinhard has
shown, it appears that a surface symmetry energy term in certain
nonrelativistic (Skyrme interaction) nuclear models can be ``dialed"
somewhat to change the neutron size without significantly spoiling the
basic fits.  Including data which are more sensitive to neutron
properties, such as isotopic trends in ground state properties, and
perhaps giant resonance energies and sum rules, could be useful to
constrain such terms.

There do exist experiments which are sensitive to neutron radii, e.g.
$\pi^+/\pi^-$ scattering, and medium energy polarized proton
scattering.  If the quoted errors on the latter can be taken literally,
one could use it to discriminate among the various models and provide
the confidence one needs to extract the desired electroweak parameters
from atomic experiments.  It would be valuable to repeat the
experiments and analyses at other energies in order to demonstrate the
consistency of the results, and to consider both $\pi^\pm$ and $\vec p$
scattering on multiple Pb isotopes for a direct experimental measure of
the isotopic shift in neutron radii.  We have also noted in this work
that the detailed distribution of neutrons, beyond just the RMS radius,
is of some importance. This implies that we may still have to rely on
the nuclear models for an extraction of the electroweak parameters.  As
discussed earlier, the use of alternative electroweak probes, such as
parity violating (polarized) electron scattering at intermediate
energies,\cite{donn} would be of obvious value for
independently extracting the desired neutron distribution.

We can turn the problem around, however, and note that an accurate
measurement of $\bar x$ from high energy experiments presents a
unique opportunity to extract the isotopic neutron
radius shifts from atomic experiments cleanly, and hence test the
nuclear models.  The situation is quite analogous to the extraction of
changes in charge radii from atomic isotope shifts.


\nonum
\section{ACKNOWLEDGMENTS}

We are grateful to C. Chinn, R. Furnstahl, N. Van Giai, M. Girod, I.B.
Khriplovich, J.  Martorell, M. Musolf, E. Ormand, P.-G. Reinhard,  P.
Ring, B.  Serot, and D. Sprung for valuable discussions and, in some
cases, the sharing of calculational data.  This work is supported in
part by U. S.  Department of Energy grants DOE/ER-06-91ER40561 and
DOE/ER-DE-FG06-88ER40427, and by NSF Grant PHY 8922274.


\figure{
 Root mean square neutron radius, in fermis, for lead isotopes, plotted
versus atomic weight. Models are defined as in Table \ref{table1a}.  We
also show points for several  additional relativistic Hartree
parameterizations, NL1, NL06, and NL075.\cite{rei,reiII} Skyrme
calculations are connected with dashes to guide the eye. Gogny HFB is
connected with dots. Relativistic models are connected with
dot-dashes.\label{fig1a}}

\figure{
Ratio of neutron to proton radius for lead isotopes, plotted versus
atomic weight. Symbols are defined as in Table \ref{table1a} and
Fig.~\ref{fig1a}.\label{fig1b}}

\figure{
Neutron correction factor $q_n$ for lead isotopes, plotted versus
atomic weight. Symbols are defined in Table \ref{table1a} and
Fig.~\ref{fig1a}. The large error bar on the right side represents the
allowed spread corresponding to a $\pm{1\over2}$\% uncertainty in the
weak charge, as given by eqn.~(6.8). The vertical position of this
error bar is arbitrary.\label{fig2}}

\figure{
Change in $q_n$ between $^{202}$Pb and
$^{208}$Pb, shown versus model weight. (The x axis is arbitrary) The
error bar on the right side represents the allowed spread in the
plotted quantity corresponding to a $\pm$1\% uncertainty in
$\bar x$, as discussed following eqn.~(6.11). The vertical
position of this error bar is again arbitrary.\label{fig3}}

\newpage
\widetext
\begin{table}
\caption{Some  properties of $^{202}$Pb relevant to
atomic parity violation, for several nuclear models.  Properties listed
are r.m.s proton radius, neutron radius, ratio $(R_n/R_p)$, difference
in mean-square charge radius from $^{208}$Pb,  $q_p$ and $q_n$, defined
in eqns.~(2.9a and b), the normalization factor, ${\cal N}$, and a
``renormalized norm'' defined in (7.1).  All distances are in fm.  The
models listed are Hartree-Fock-Bogolyubov with a Gogny\cite{gir}
finite-ranged D1S interaction, using a 15 shell spherical harmonic
oscillator basis (G:HFB), various parameterizations of the Skyrme
interaction\cite{gar,fri} in the spherical Hartree-Fock
approximation, Skyrme A (SkA), star (Sk*), and  3 (Sk3), and
relativistic Hartree mean field calculations \cite{rei,reiII} with
a nonlinear PL40 parameter set (Rel).  The first row contains
experimental numbers, where known.\cite{auf,jag,tho} The
final row simply indicates the maximum spread among the models.}
\bigskip
\begin{tabular}{ccccccccc}
 $^{202}$Pb
	& $\sqrt{r_p^2}$
        	& $\sqrt{r_n^2}$
			& $R_n/R_p$
				& $\delta\,r_{ch}^2$
						& $q_p$  & $q_n$  &
				                ${\cal N}$ & ${\cal N}' $ \\
\tableline
\ Data\ &      &	&	& -.330(4)&     &        &        &        \\
G:HFB	& 5.409& 5.519	& 1.020	&-.274 & .90599 & .90318 & 1.0087 & 1.0041 \\
SkA	& 5.431& 5.607  & 1.032	&-.297 & .90604 & .90141 & 1.0069 & 1.0039 \\
Sk*	& 5.423& 5.560	& 1.025	&-.296 & .90608 & .90254 & 1.0076 & 1.0039 \\
Sk3	& 5.488& 5.590	& 1.019	&-.355 & .90638 & .90382 & 1.0018 & 1.0028 \\
Rel	& 5.484& 5.742	& 1.047	&-.319 & .90618 & .89949 & 1.0023 & 1.0034 \\
\tableline
Spread$\sim$
        & .08	& .2  	& .03	& .08 & .0004  & .004  & 0.007 & .001 \\
\end{tabular}
\label{table1a}
\end{table}
\bigskip
\widetext
\begin{table}
\caption{
 Same as Table I,  for $^{208}$Pb.}
\bigskip
\begin{tabular}{ccccccccc}
 $^{208}$Pb
	& $\sqrt{r_p^2}$
        	& $\sqrt{r_n^2}$
			& $R_n/R_p$
				& $\delta\,r_{ch}^2$
						& $q_p$  & $q_n$  &
				                ${\cal N}$ & ${\cal N}' $ \\
\tableline
\ Data\ & 5.453(2)& 5.59(4)&1.03&0.0  & .906(1)&        & $\equiv$1.0 &  \\
G:HFB	& 5.435 & 5.569&1.025& 0.00   & .90596 & .90260 & 1.0068 & 1.0018 \\
SkA	& 5.459 & 5.670&1.039& 0.00   & .90599 & .90051 & 1.0049 & 1.0016 \\
Sk*	& 5.451 & 5.620&1.031& 0.00   & .90605 & .90176 & 1.0053 & 1.0016 \\
Sk3	& 5.521 & 5.646&1.023& 0.00   & .90636 & .90334 & 0.9992 & 1.0005 \\
Rel	& 5.513 & 5.822&1.056& 0.00   & .90607 & .89813 & 1.0003 & 1.0012 \\
\tableline
Spread$\sim$
        & .1   & .3    & .03 &        & .0004  & .005   & 0.008 & .001 \\
\end{tabular}
\label{table1b}
\end{table}
\bigskip
\mediumtext
\begin{table}
\caption{ Changes in various nuclear properties between
$^{202}$Pb and $^{208}$Pb. The models are the same as in Table 1.
Properties listed are the change in relative neutron to proton radii,
the change in proton and neutron correction factors $q_n$ and $q_p$,
and the relative change in the norm. The final column is for the
``renormalized norm" defined in eqn. (7.1).  The last row shows the
maximum  spread among the models.}
\bigskip
\begin{tabular}{cccccc}
$^{202}$Pb $\rightarrow\ ^{208}$Pb
        & $\Delta R_n/R_p$
		& $\Delta q_p$
				   & $\Delta q_n$
				&$\Delta{\cal N}$& $\Delta{\cal N'}$ \\
\tableline
G:HFB	& .0043 & 3$\cdot 10^{-5}$ & 5.8$\cdot 10^{-4}$
				& .0020		& .00227 \\
SkA	& .0064 & 5$\cdot 10^{-5}$ & 9.0$\cdot 10^{-4}$
				& .0021		& .00232 \\
Sk*	& .0056 & 3$\cdot 10^{-5}$ & 7.8$\cdot 10^{-4}$
				& .0022 	& .00225 \\
Sk3	& .0040 & 2$\cdot 10^{-5}$ & 4.8$\cdot 10^{-4}$
 				& .0026 	& .00230 \\
Rel	& .0091 & 1$\cdot 10^{-4}$ & 1.4$\cdot 10^{-3}$
 				& .0021		& .00218 \\
\tableline
Spread$\sim$
       & .0051 & 8$\cdot 10^{-5}$  &  9$\cdot 10^{-4}$
				&  6$\cdot 10^{-4}$ & 1$\cdot 10^{-4}$\\
\end{tabular}
\label{table2}
\end{table}

\end{document}